\documentstyle[graphics]{aa}
\newcommand{\ls}
 {\mathrel{\hbox{\rlap{\hbox{\lower4pt\hbox{$\sim$}}}\hbox{$<$}}}}
\newcommand{\gs}
 {\mathrel{\hbox{\rlap{\hbox{\lower4pt\hbox{$\sim$}}}\hbox{$>$}}}}
\newcommand{\degg}{\hbox{$^\circ$}}
\newcommand{\arcm}{\hbox{$^\prime$}}
\newcommand{\arcs}{\hbox{$^{\prime\prime}$}}
\newcommand{\et}{{\it et al.}\ }

\newcommand{\rosat}{{\it ROSAT}}

\newcommand{\asca}{{\it ASCA}}

\newcommand{\xmm}{{\it XMM-Newton}}

\def\la{\mathrel{\hbox{\rlap{\hbox{\lower4pt\hbox{$\sim$}}}{\raise2pt\hbox{$
<$}}
}}}
\def\ga{\mathrel{\hbox{\rlap{\hbox{\lower4pt\hbox{$\sim$}}}{\raise2pt\hbox{$
>$}}
}}}

\begin{document}

%\thesaurus{(11.01.2; 11.17.4(PKS 0537-286); 13.25.2)}

\title{The first XMM-Newton spectrum of a high redshift quasar - 
PKS~0537$-$286.}
\author{J.N. Reeves\inst{1} 
\and M.J.L.\ Turner\inst{1}
\and P.J\ Bennie\inst{1}
\and K.A.\ Pounds\inst{1}
\and A.\ Short\inst{1}
\and P.T. O'Brien\inst{1}
\and Th. Boller\inst{2}
\and M. Kuster\inst{3}
\and A. Tiengo\inst{4}}

\offprints{J.N. Reeves}

\institute{X-Ray Astronomy Group; Department of Physics and Astronomy;
Leicester University; Leicester LE1 7RH; U.K.
\and Max-Planck-Institut f{\"u}r extraterrestrische Physik, Postfach 1603,
85748 Garching, Germany
\and Institut f{\"u}r Astronomie und Astrophysik - Astronomie, University
of T{\"u}bingen, Waldh{\"a}user Strasse 64, D-72076 T{\"u}bingen, Germany
\and Vilspa, Villafranca Satellite Tracking Station, Apartado 50727,
28080 Madrid, Spain}

\date{Received September 2000 / Accepted October 2000}

\maketitle

\begin{abstract}

We present \xmm\ observations of the high redshift ($z=3.104$), 
radio-loud quasar PKS~0537$-$286. The EPIC CCD 
cameras provide the highest signal-to-noise spectrum of a high-z 
quasar to date. The EPIC observations show that PKS~0537$-$286 is 
extremely X-ray luminous (L$_{X}=2\times10^{47}$~erg~s$^{-1}$), with an 
unusually hard X-ray spectrum ($\Gamma=1.27\pm0.02$). 
The flat power-law emission extends over the whole observed energy range 
(0.4 to 40 keV in the quasar rest frame); 
there is no evidence of intrinsic absorption, which 
has been claimed in PKS~0537$-$286 and other high $z$ quasars. 

However, there is evidence for weak Compton reflection. 
A {\it redshifted} iron K line, observed at 1.5~keV - 
corresponding to $\sim$6.15~keV in the quasar rest frame - 
is detected at 95\% confidence. If confirmed, this is the most distant
iron K line known. 
%The centroid energy may even imply that
%the line originates from the inner quasar accretion disk,
%nearest to the putative super-massive black hole. 
The line equivalent width is small (33eV), consistent 
with the `X-ray Baldwin effect' observed in other luminous quasars. 
The reflected continuum is also weak (R~$\la$0.25). 
We find the overall spectral 
energy distribution of PKS~0537$-$286 is dominated by the 
X-ray emission, which, together with the flat power-law and weak 
reflection features, suggests that the X-radiation from PKS~0537$-$286 
is dominated by inverse Compton emission associated with a 
face-on relativistic jet.


\end{abstract}

\section{Introduction}

At the present time there is little known about the X-ray spectra
of high redshift quasars, and thus a lack of understanding of the nature
of the quasar central engine at the highest redshifts. However the
situation is improving; the \rosat\ 
mission substantially increased the number of high $z$ quasars with
detected X-ray emission (Brinkmann \et 1997), whilst \asca\
obtained the first broad-band spectra 
(e.g. Cappi \et 1997, Reeves \et 1997, Vignali \et
1999), albeit with rather limited photon statistics and sensitivity.  
The high throughput and bandpass of \xmm\ is about to revolutionize 
studies of high redshift quasars. Many high redshift quasars should be
found routinely through follow-up surveys of XMM fields (Watson
\et 2001), detailed spectral studies can be performed on the 
brightest objects
and the high sensitivity limit of \xmm\ should enable the detection of
the most distant known quasars at $z>5$ (Brandt \et 2000).

At $z=3.104$, PKS~0537$-$286 is one of the brightest known high $z$ quasars, 
first recognised as a 1 Jy Parkes southern radio source 
(e.g. Bolton \et 1975) and later identified as a QSO (Wright \et 1978). 
The optical (rest frame UV) spectrum shows two strong absorption
systems associated with the quasar, one of which
produces a cut-off in the quasar continuum at the
Lyman-limit. The redshift of PKS~0537$-$286 has been confirmed at 
$z=3.104\pm0.001$ (Osmer \et 1994). PKS~0537$-$286 was first detected 
as an X-ray source by the Einstein observatory
(Zamorani \et 1981), with a position within 1\arcm\ of the radio and
optical source. The identification was later confirmed by \rosat\
(B\"{u}hler \et 1995), and with \asca\ there were enough counts
to obtain a crude X-ray spectrum (Siebert \et 1996, Cappi \et 1997,
Reeves \et 1997). These observations revealed a luminous X-ray source, with
a flat X-ray spectrum perhaps indicative of a soft X-ray cut-off due to
absorption. We now present the \xmm\ observations of PKS~0537$-$286. 
Values of $ H_0 = 50 $~km\,s$^{-1}$\,Mpc$^{-1}$ and $ q_0 = 0.5 $
have been assumed and all fit parameters are given in the quasar rest-frame.

%\vspace{-0.5cm}
\section{XMM-Newton observations}

PKS~0537$-$286 was observed during orbit 51 of the Cal-PV phase of \xmm. 
Processing was performed using the XMM SAS (Science Analysis Software). 
The initial data files were processed with the \textsc{emchain} 
and \textsc{epchain} scripts, using the latest known calibration. 
X-ray events corresponding to patterns 0-12 for the 2 MOS cameras 
(similar to grades 0-4 in \asca) were used; for the PN, only 
pattern 0 events (single pixel events) were selected. For the MOS 
detectors, electronic noise was screened out 
by rejecting events with negative E3 values. Known hot or bad 
pixels were also removed during screening. We refer the reader to
Turner \et (2001) for an in-depth description of the EPIC-MOS
instruments, whilst a detailed account of the EPIC-PN detector 
can be found in Str\"{u}der \et (2001) . 
The non X-ray background remained relatively low
throughout the observations, so it was not necessary to filter 
events on time. We also found no variability in the
source count rate during the observation.

The screening process yielded 19.2 ksec of data for MOS-1, 
28.0 ksec for MOS-2 and 38 ksec for the PN detector. Source spectra 
were extracted from circular regions of 30\arcs\ radius for the PN
(to avoid the edge of the chip) and 1\arcm\ radius for the MOS.  
Background spectra were taken from an 
identical circular region, offset from the source position. 
The majority of source counts fall onto these source regions 
(at least 80\% for the PN and $>$90\% for the MOS), whilst the
background count rates are very low; $1.07\pm0.17\times10^{-2}$~cts~s$^{-1}$,
$4.8\pm1.5\times10^{-3}$~cts~s$^{-1}$ 
and $8.8\pm1.1\times10^{-3}$~cts~s$^{-1}$ for the PN, MOS-1 and MOS-2
respectively. The net source count rates obtained are 
0.257$\pm$0.004~cts/s, 0.268$\pm$0.003~cts/s and
0.513$\pm$0.004~cts/s, for MOS-1, MOS-2 and the PN respectively. 
At this level, even in full-frame mode, photon pile-up 
is negligible. We do not consider the RGS data in this paper, 
as the count rate is too low ($<0.1$~cts/s) for a reliable analysis 
of the dispersed spectrum. The background subtracted EPIC spectra 
were fitted, using \textsc{xspec v11.0}, with the latest response
matrices produced by the EPIC team; the systematic level of uncertainty 
is $<5$\%.  Spectra were binned to a minimum of 20 counts per
bin, to apply the $\chi^2$ minimisation technique. All 
subsequent errors are quoted at 90\% confidence 
($\Delta\chi^2=4.6$ for 2 interesting parameters). 

\begin{figure}
\resizebox{\hsize}{!}{\rotatebox{-90}{\includegraphics{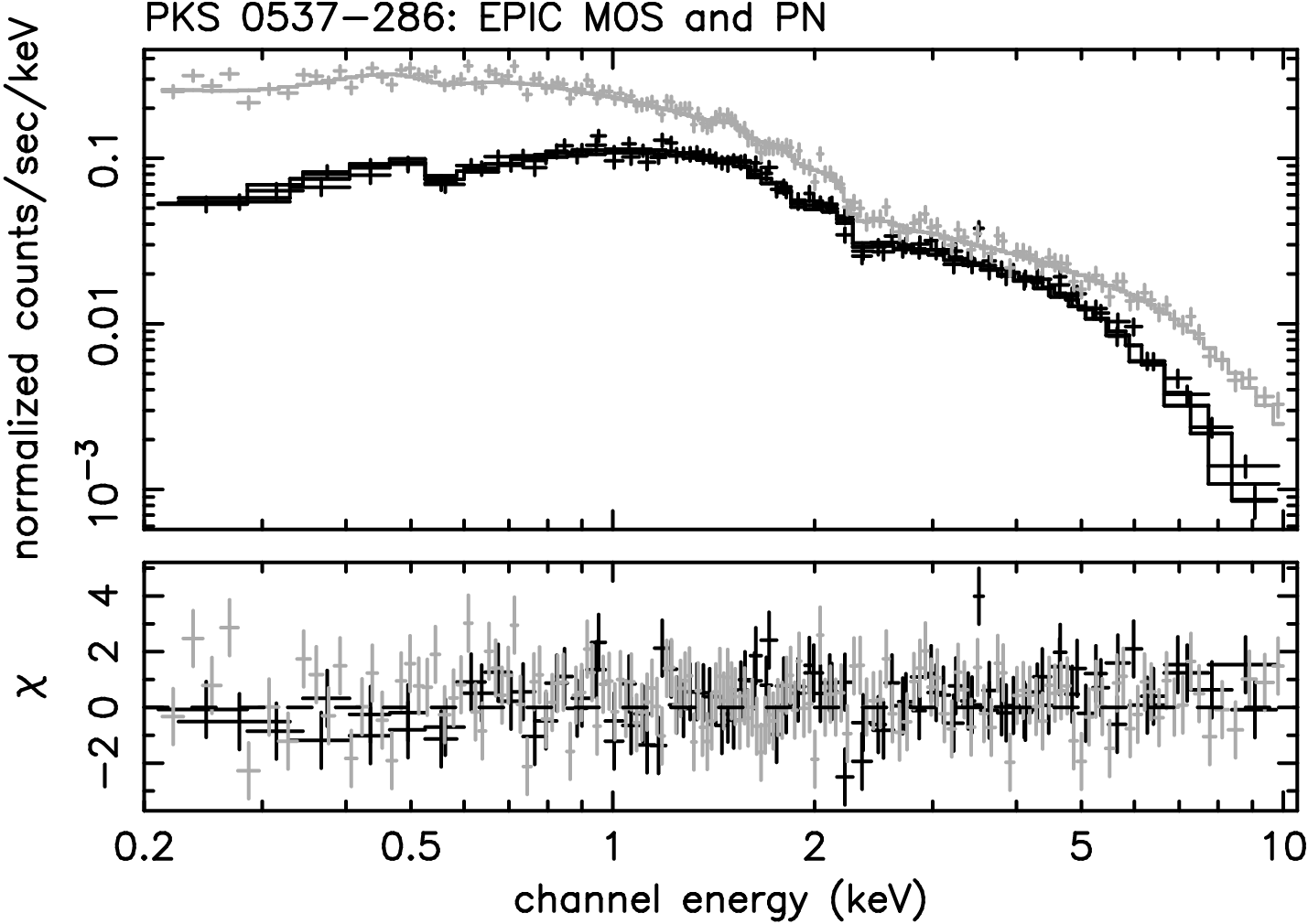}}}
\caption{EPIC-MOS and EPIC-PN (greyscale) spectra of the quasar PKS
0537-286. The spectrum corresponds to fit 2 in table 1 (power-law with
iron line and K edge). 
The flat ($\Gamma=1.27$) power-law extends over the whole QSO
energy range, from 0.8 - 40 keV, with no sign of a low energy cut-off.} 
\end{figure}

%\vspace{-0.5cm}
\section{EPIC Spectral Analysis}

\begin{figure}
\resizebox{\hsize}{!}{\rotatebox{-90}{\includegraphics{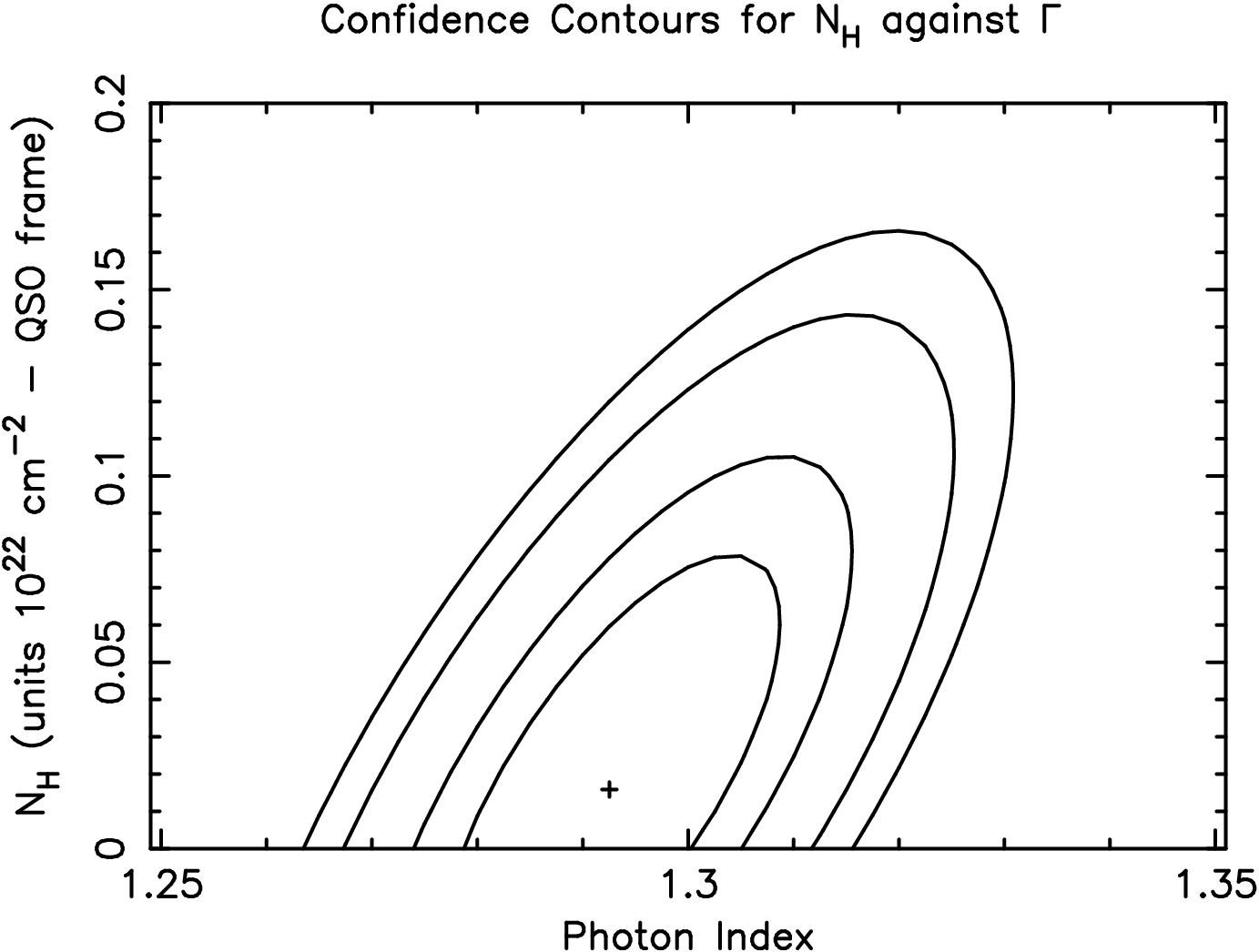}}}
\caption{A confidence contour plot, showing the quasar photon index
against the column density. Contours represent the 68\%, 90\%, 99\%
and 99.9\% 
levels. There is no X-ray absorption in PKS
0537-286 down to a limit of $1\times10^{21}$~cm$^{-2}$ in the QSO
frame (at 90\% confidence).}
\end{figure}

We initially fitted the individual X-ray spectra from the 3 EPIC 
detectors, to check for consistency between the datasets. From 
a simple power-law fit it was found that the photon index 
agreed for all 3 detectors ($\Gamma$=1.25-1.30) and that the relative 
normalisations were consistent to within 5\%. 
Therefore we proceeded to fit all three datasets simultaneously, 
assuming a Galactic column of $N_{H}=1.95\times10^{20}\rm{cm}^{-2}$ 
(Elvis \et 1989). The EPIC spectrum is plotted in Figure 1 and  
the results from a power-law fit to the broad-band data are shown as fit 1 
(in table 1). Clearly a power-law is a good fit ($\chi_{\nu}^2\sim1$),
with no obvious continuum curvature present. We note that the power-law 
photon index of $\Gamma=1.27\pm0.02$ is considerably flatter than the 
mean quasar index (e.g. Reeves \& Turner 2000), even for 
radio-loud quasars which have a typical mean of $\Gamma=1.6$. 
The observed broad-band (0.2-10 keV) flux is 
3.1$\times10^{-12}$~erg~cm$^{-2}$~s$^{-1}$, which, under the assumption of 
isotropic X-ray emission, corresponds to a luminosity at $z=3.104$ 
of 2.2$\times10^{47}$ erg~s$^{-1}$ (0.8-40 keV in the QSO frame). 
This makes PKS~0537$-$286 one of the most luminous known quasars in 
the Universe. 

We also searched the spectrum for signs of photoelectric absorption. 
There are several reported cases in the literature of excess X-ray 
absorption towards high z quasars, from \rosat\ and \asca\
observations (see section 4). 
The excellent soft X-ray response of the EPIC instruments allows us 
to place very tight limits on the amount of obscuring material.  
However we found no excess absorption 
towards PKS~0537$-$286, the power-law extending un-broken down to 0.2 keV. 
A confidence contour plot is shown as Figure 2; we place a 90\% 
upper limit on the intrinsic column of $N_{H}<5.0\times10^{19}\rm{cm}^{-2}$, 
which in the quasar rest frame (at z=3.104) corresponds to 
$N_{H}<1.0\times10^{21}\rm{cm}^{-2}$.   

\begin{table*}
\centering
\caption{X-ray spectral fits to PKS~0537$-$286. $^a$ Rest energy of the iron
K line or edge in keV (QSO frame). $^b$ Equivalent width of the line in
eV. $^c$ Optical depth of the iron K edge. $^d$ Strength of the reflection
component R (=$\Omega/2\pi$). $^e$ Assumed value for high energy
cut-off in keV. $^f$ Indicates parameter is fixed.}

\begin{tabular}{@{}llcccccccc@{}}
\hline                 
Fit & Model & $\Gamma$ & \multicolumn{2}{c}{Iron line} &
\multicolumn{2}{c}{Iron edge} & \multicolumn{2}{c}{Reflection} & 
$\chi^{2}$/dof \\   

\ & \ & \ & E$^a$ & EW$^b$ & E$^a$ & $\tau^c$ & R$^d$ & E$_{c}$$^e$ \\

\hline

1. & PL only  & 1.27$\pm0.02$ & & & & & & & 1049.2/1000 \\

2. & PL + GA + EDGE & 1.27$\pm$0.02 & 6.15$\pm$0.15 & 33$^{+30}_{-23}$ & 
7.1$^f$ & 0.05$^{+0.05}_{-0.04}$ & & & 1038.5/997 \\

3. & PL + GA + PEXRAV & 1.28$\pm$0.04 & 6.15$^f$ & 33$^f$ & & &
0.25$^{+0.11}_{-0.09}$ & 100$^f$ & 1044.5/998 \\

4. & PL + GA + PEXRAV & 1.29$\pm$0.04 & 6.15$^f$ & 33$^f$ & & &
0.11$^{+0.09}_{-0.08}$ & 250$^f$ & 1038.6/998 \\

5. & PL + GA + PEXRAV & 1.30$\pm$0.04 & 6.15$^f$ & 33$^f$ & & &
0.03$^{+0.07}_{-0.03}$ & none & 1036.9/998 \\

\hline
\end{tabular}
\end{table*}

\subsection{The iron K emission properties} 

\begin{figure}
\resizebox{\hsize}{!}{\rotatebox{-90}{\includegraphics{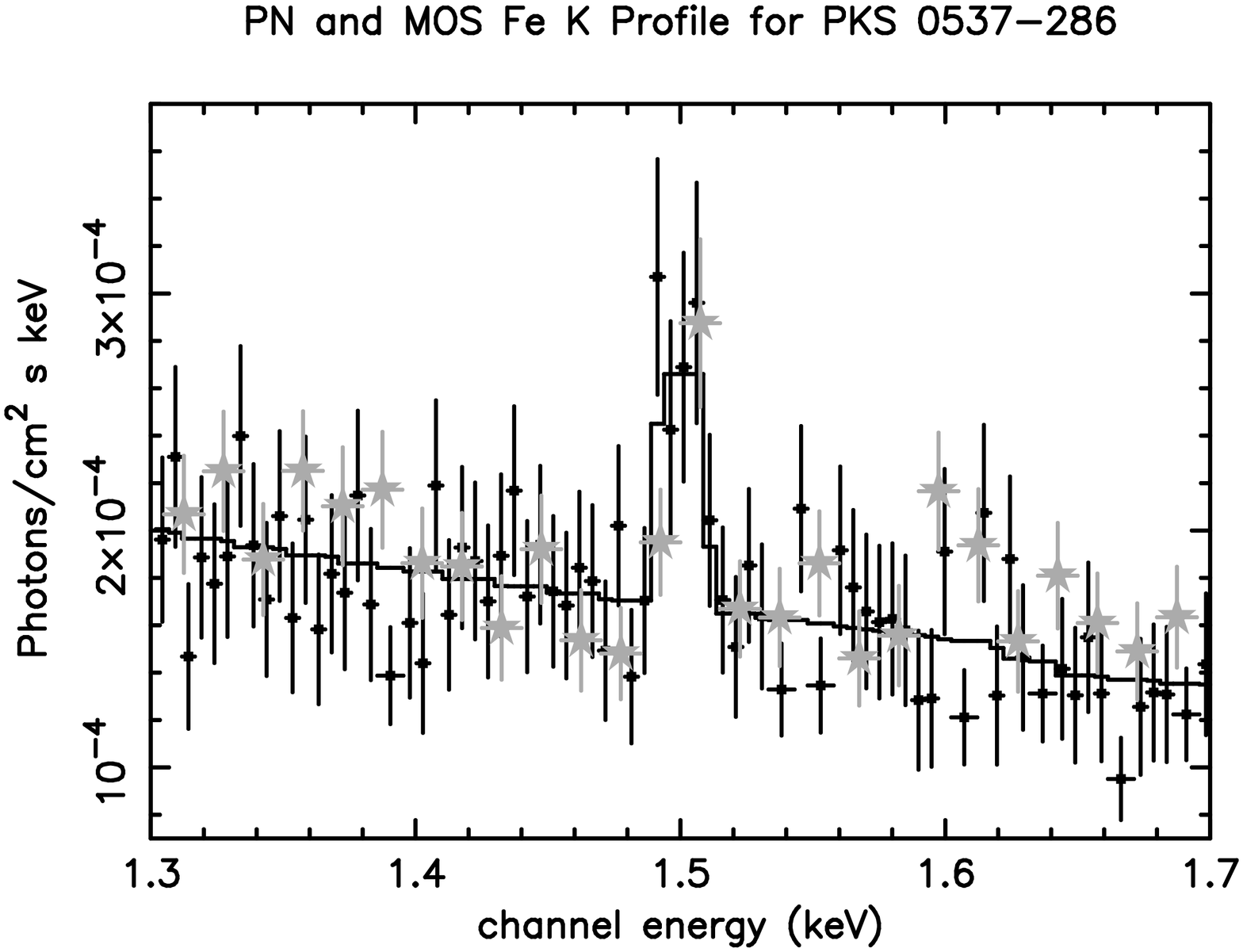}}}
\caption{A close-up of the EPIC PN and MOS (greyscale) data 
between 1.3-1.7 keV. The iron K emission line, at $z=3.104$, is
redshifted down to 1.5 keV in the observed spectrum. The MOS-1 and
MOS-2 data have been co-added.} 
\end{figure}

The relative brightness of PKS~0537$-$286 allows us, for the first 
time, to search for signs of the iron K$\alpha$ emission line in a 
high redshift quasar. We restricted our search to the 1-2 keV 
band, where the iron K line should appear at this redshift; conveniently 
the throughput of the \xmm\ EPIC instrument reaches a maximum in 
this band. We initially fixed the line width at $\sigma=10$~eV and  
the result is shown as fit 2. 
A narrow Fe K line is 
detected (equivalent width EW=33eV, QSO frame), at 95\% confidence 
for 2 parameters, whilst the line peaks at 6.15$\pm$0.15~keV (or 1.5 keV 
observed-frame). A close up of the EPIC PN and MOS data, illustrating the line
detection at $z=3.104$, is shown in Figure 3.
The confidence contours to the line, from a fit to the MOS and PN
data, are also shown in Figure 4. 
Although the line energy is consistent with neutral material, the
apparent redshift of the line may indicate an origin from the inner
quasar accretion disk (Fabian \et 1989) or from
reprocessing in infalling matter.  
The velocity width of the line is not well constrained, with an upper 
limit of $\sigma<500$~eV. The line emission is considerably weaker
than the Fe lines observed in Seyfert~1 
galaxies (Nandra \et 1997) and in some radio-quiet quasars 
(Reeves \& Turner 2000). 

\begin{figure}
\resizebox{\hsize}{!}{\rotatebox{-90}{\includegraphics{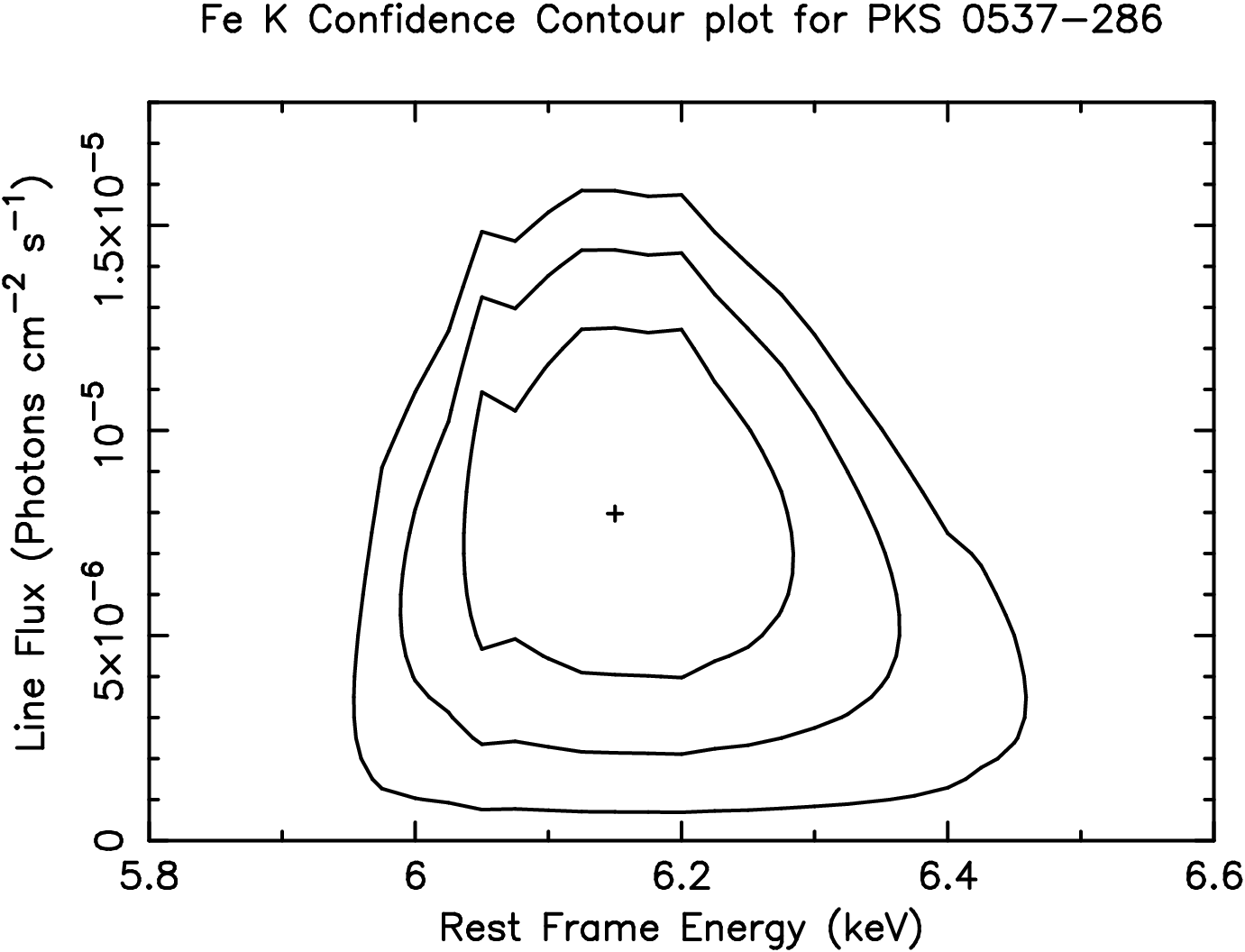}}}
\caption{Confidence contours for the iron K line in PKS~0537$-$286. 
Contours represent the 68\%, 90\% and 95\% levels. The Fe K line is
detected at 95\% confidence. Note the slight redshift of the line
energy.}
\end{figure}

Finally we tried to constrain the iron K absorption edge, which in 
disk reflection models (e.g. George \& Fabian 1991) is predicted to 
accompany the iron K line. We fixed the edge energy at 7.1 keV,
consistent with neutral iron. We obtained a value of
$\tau=0.05^{+0.05}_{-0.04}$, although the edge is only detected at
90\% confidence (see fit 2). 
 
\subsection{The effect of Aluminium K$\alpha$ on the Iron Line}

Since the proposed iron K line (observed at 1.5 keV) lies 
near the Al-K$\alpha$ line, at 1.487 keV, in the detector
background, we have checked to see whether this can
contribute towards the observed Fe line flux at 1.5 keV. To do this we
extracted a large source-free background region, of 4\arcmin\
radius, from the PKS~0537$-$286 PN field, and then measured the
strength of the Al K$\alpha$ line directly. We re-normalised the
background fluxes (from an area 63 times larger), to the size
of our source extraction region. It was found that the total integrated
count rate from the Al K$\alpha$ line was 
$7.5\pm2.6\times10^{-5}$~ct~s$^{-1}$, compared to the quasar line count
rate of $1.85\pm0.6\times10^{-3}$~ct~s$^{-1}$. Thus we can 
conclude that that the Al K$\alpha$ line contributes only $\sim$4\% of the
quasar line flux, {\it before background subtraction}. Note that at
1.5 keV, the total background count rate is $<$1\% of the total quasar count
rate. There are also no strong variations in background, across the
central chips, for either the MOS or PN. 

As a final check, we also used a long closed-filter PN observation 
(in orbit 59) to estimate the background count rate from the 
Al K$\alpha$ line. We found that the total flux in the Al line  
(for an area the size of the PN source extraction region) was 
$9.0\times10^{-5}$~ct~s$^{-1}$, consistent with the value given above,
corresponding to only 5\% of the quasar line counts. Therefore 
it is highly unlikely that the background subtraction at 1.5~keV contributes
towards the detection of the quasar iron K line.

\subsection{Constraints on the Compton Reflection Continuum}

Having detected these weak iron K features, we next tried to 
constrain the properties of the Compton reflection `hump', 
particularly as the {\it observed} emission extends up to 40 keV in the 
quasar rest frame. We used the \textsc{pexrav} model in XSPEC 
(Magdziarz \& Zdziarski 1995), assuming a disk inclination of 30\degg 
and a high energy cut-off equal to 100~keV. We left R, the fraction of
material (of $2\pi$~steradian solid angle) that is irradiated by the 
X-ray source (where R=$\Omega/2\pi$), as a free parameter in the fits. 
The results are given as fit 3. 
We find a best-fit value for the reflection continuum of 
R$=0.25^{+0.11}_{-0.09}$, indicating that the reflecting material 
subtends a solid-angle lower than the 2$\pi$ steradians expected 
from an accretion disk geometry. The small amount of
reflection (with R$\sim$0.25), is consistent with the low equivalent
width of the iron K line. 

We note however, that our assumption for the high energy 
cut-off of the hard X-ray power-law may effect the fits. To test this 
we tried two additional fits, one with a cut-off at 250~keV and one 
fit with no high energy cut-off (fits 4-5). 
For the former we find that the reflection component is even 
weaker (R$=0.11^{+0.09}_{-0.08}$), whilst with no cut-off the value 
of R is consistent with zero (R~$<0.10$). Comparing the 
statistics of fits 3-5, there is a preference for very weak 
reflection and no high energy cut-off, although the differences 
are not significant. Finally we tried fixing R at 1 in our fits. 
In this case, whatever cut-off energy was 
assumed, the fit obtained was always considerably worse 
(by $\Delta\chi^2>100$, or at $>99.99$\% confidence). 
 
\subsection{The Spectral Energy Distribution of PKS~0537$-$286}

In Figure 5, we compile all the published multi-wavelength data on 
PKS~0537$-$286, together with the \xmm\ spectrum. The SED (spectral 
energy distribution) of PKS~0537$-$286 is extreme when compared to a 
typical radio-loud quasar (Elvis \et 1994). In PKS~0537$-$286 the 
majority of the observed quasar power is radiated in the X-ray band 
(typically in AGN the X-ray emission is only 5\% of the bolometric 
luminosity; e.g. Elvis \et 1994). The optical-UV  
emission is also relatively weak in comparison to the X-rays 
($\alpha_{ox}=-0.94$). However this is consistent with the
observations of other high redshift blazar-like AGN (see Fabian \et
1999), where a strong, but flat, X-ray emission component is also
seen.  

\section{Discussion}

The \xmm\ observations have showed that PKS~0537$-$286 is an 
extremely luminous quasar, with a particularly flat X-ray
spectrum and weak iron K emission and reflection features. 
Indeed the total power output of PKS~0537$-$286 is dominated by the 
X-ray emission. A likely interpretation is 
that the X-rays mainly arise through the inverse Compton 
emission associated with a relativistic jet. In this scenario 
relativistic electrons (or positrons) in the jet  
Compton upscatter either the low energy synchrotron photons 
produced in the jet (Synchrotron Self Compton - Jones,
O'Dell \& Stein 1974, Ghisellini, Maraschi \& Treves 1985) or 
alternatively the thermal optical/UV photons from the accretion 
disk (External Inverse Compton - Ghisellini \& Madau 1996). 
This can naturally account for the hard spectrum of PKS~0537$-$286, 
as in the X-ray band we are observing the onset
of this Inverse Compton component, presumably extending out to
$\gamma$-ray energies. 

\begin{figure}
\resizebox{\hsize}{!}{\rotatebox{-90}{\includegraphics{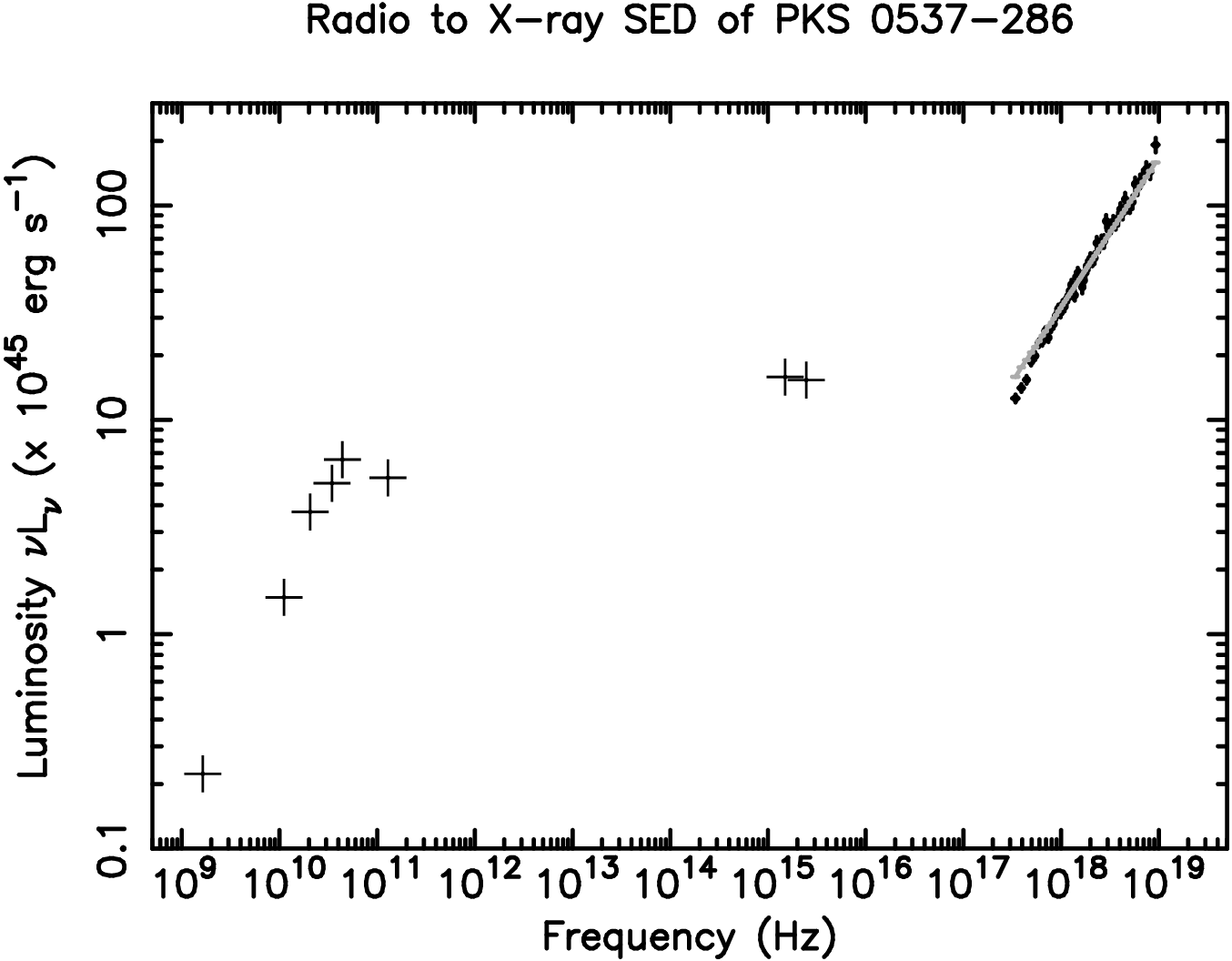}}}
\caption{The radio to X-ray spectral energy distribution of
PKS~0537$-$286. The total power output of the quasar is dominated by the
X-ray emission. The SED is similar to other high luminosity 
blazars at high redshift (see Fabian \et 1999).}
\end{figure}

As it appears that we can rule out a scenario whereby all
of the primary X-ray emission originates above the accretion disk, 
we instead tried to fit the data with 
two continuum emission components; a `Seyfert-like' steep power-law 
(with $\Gamma$ fixed at 1.9) originating from near the
accretion disk, and a hard, bright jet component with a flat X-ray
slope (i.e. $\Gamma=1.2-1.3$ as before). We modified the Seyfert-like 
power-law with a continuum reflection component
(assuming R=1) and also added a narrow iron K line of equivalent width
150~eV. Although poorly constrained, we obtained a good fit to the
data when the Seyfert-like component was 
typically 4-5 times weaker than the flat jet component. 
Thus in PKS~0537$-$286 we may be observing two X-ray components; the
steep Seyfert-like emission from near the disk, responsible for the iron K
line, and the hard component from the jet, which considerably dilutes
the disk reflection and line features. 

This provides a possible explanation for the apparent 
weakness of the iron K line and Compton reflection hump, when 
compared to other observations of Seyfert galaxies and quasars.  
If the relativistic jet is orientated close to the line-of-sight, as 
expected in unification schemes for quasars (Antonucci 1993), then 
any disk reflection features can be diluted due to Doppler 
boosting of the featureless X-ray continuum, through the bulk motion 
of the jet. This is consistent with other observations of X-ray
luminous, predominantly radio-loud quasars, where the strength of the
iron line is diminished (the so-called `X-ray Baldwin effect' - 
Iwasawa \& Taniguchi 1993). 
For PKS~0537$-$286, at a redshift 
of $z=3.104$, any iron K emission features are redshifted into a more 
favourable part of the EPIC bandpass, where the effective area is near
its maximum. {\it This has enabled us to detect a weak iron K line in  
PKS~0537$-$286; indeed at $z=3.104$ this is the most distant 
iron K emission line detected to date.} 
The limited statistics do not allow the line width or shape to be
constrained. However, the median energy of 6.15~keV implies
reprocessing in `cold' matter close to the central black hole or from
infalling material at $v\sim13000$~km~s$^{-1}$. 
Clearly, more detailed studies with \xmm\ of other high redshift quasars, both
radio-loud and radio-quiet, are required to explore this.  

Past studies with \rosat\ and \asca\ have showed that several 
high redshift radio-loud quasars appear to show 
intrinsic X-ray absorption, above that from our own Galaxy (see 
Bechtold \et 1994, Elvis \et 1994b, Reeves \et 1997, Cappi \et 1997, 
Fiore \et 1998). Correlations involving samples of quasars 
(Reeves \et 1997, Reeves \& Turner 2000) have showed that this 
apparent excess absorption is correlated with object redshift, with 
many high z quasars requiring substantial absorption columns (up to 
$10^{22}$~cm$^{-2}$), whilst low z quasars remain unabsorbed. 
Recently even, evidence has been presented for substantial soft 
X-ray absorption towards one blazar-like object (GB~1428+4217) at 
$z=4.71$ (Boller \et 2000). At present the nature of this 
absorption is unclear, but it could be related to the dense 
Galactic-scale environments proposed in the early evolutionary 
stages of quasars (see Fabian 1999). Alternatively in some objects the
X-ray absorption may originate from intervening damped Lyman-$\alpha$ systems,
which are thought to be associated with the disks of young protogalaxies 
(e.g. Wolfe \et 1994).  

From the published optical (rest frame UV) spectrum, it is known 
that PKS~0537$-$286 has substantial UV absorption, with at least two 
different Lyman-limit systems observed close to the known redshift of 
the quasar (Wright \et 1978, Osmer \et 1994). The velocities of these 
systems are $\Delta\nu=4200$ 
and 40000~km~s$^{-1}$, {\it blueshifted} with respect to the quasar. The 
latter system has a very deep Lyman-continuum cut-off at 912\.{A}
(rest frame), where the quasar continuum flux drops close to zero. 
So it perhaps
seems surprising at first that in this \xmm\ observation we detect 
no intrinsic X-ray absorption towards PKS~0537$-$286. However in order
to measure appreciable line-of-sight X-ray absorption 
(of the order $N_{H}\sim10^{22}$~cm$^{-2}$), one would have to
encounter a damped Lyman-$\alpha$ system. 
We can now rule this scenario out in PKS~0537$-$286, and note the chance
coincidence of such a system out to $z=3$ is $\sim$20-30\% (e.g. Steidel
1992).  

The \xmm\ observation is in contrast to some earlier 
\asca\ and \rosat\ observations of PKS~0537$-$286, where a low energy 
X-ray cut-off had been claimed (Reeves \& Turner 1997, Cappi \et 1997). 
One possibility is that the absorption is variable; we note the `warm' 
absorbers observed in low z AGN are known to vary 
(e.g. MCG-6-30-15, Otani \et 1996; NGC 3227, George \et 1998). 
It is also possible that if the absorption is associated with a high
velocity system, the absorbing material may move out of the 
line-of-sight. Finally, the photon statistics on the earlier \asca\
and \rosat\ observations were quite poor, so it is possible that
the earlier detection was spurious. We conclude that it will be 
particularly important to utilise future \xmm\ 
observations to estimate the frequency of these `absorbed quasars', 
as a probe of some of the most distant matter in the Universe,
and to determine the environments of some of the earliest
quasars. 

%\vspace{-0.5cm}
\section*{ Acknowledgements }

This paper is based on observations obtained with XMM-Newton, 
an ESA science mission with instruments and contributions directly
funded by ESA Member States and the USA (NASA).
EPIC was developed by the EPIC Consortium led by the Principal 
Investigator, Dr. M. J. L. Turner. The consortium comprises the 
following Institutes: University of Leicester, University of 
Birmingham, (UK); CEA/Saclay, IAS Orsay, CESR Toulouse, (France); 
IAAP Tuebingen, MPE Garching,(Germany); IFC Milan, ITESRE Bologna, 
OAPA Palermo, Italy. EPIC is funded by: PPARC, CEA, CNES, DLR and
ASI. 

We would like to thank the EPIC instrument team, for their hard work during
the calibration phase and the SOC and SSC teams for making the
observation and subsequent analysis possible. We also thank the
anonymous referee for his/her prompt response and helpful
suggestions.

%\vspace{-0.5cm}

\label{lastpage}

\end{document}